\documentclass[onecolumn]{aastex631}

\shorttitle{}
\shortauthors{K Anand et al. 2024}
\graphicspath{{./}{figures/}}

\begin{document}

\title{Detection of simultaneous QPO triplets in 4U 1728-34 and constraining the neutron star mass and moment of inertia}



\author{Kewal Anand}
\affiliation{Department of Physics, Indian Institute of Technology Kanpur, Kanpur Nagar, Uttar Pradesh 208016, India}
\author{Ranjeev Misra}
\affiliation{Inter-University Center for Astronomy and Astrophysics, Ganeshkhind, Pune, Maharashtra
	411007, India}
\author{J S Yadav}
\affiliation{Space, Planetary and Astronomical Sciences and Engineering, IIT Kanpur, Kanpur Nagar, Uttar Pradesh 208016, India}
\affiliation{Department of Astronomy and Atrophysics, Tata Institute of Fundamental Research, Mumbai, Maharashtra  400005, India}
\author{Pankaj Jain}
\affiliation{Space, Planetary and Astronomical Sciences and Engineering, IIT Kanpur, Kanpur Nagar, Uttar Pradesh 208016, India}
\author{Umang Kumar}
\affiliation{Department of Physics, Ashoka University,  Sonipat, Haryana 131029, India}
\author{Dipankar Bhattacharya}
\affiliation{Inter-University Center for Astronomy and Astrophysics, Ganeshkhind, Pune, Maharashtra
	 411007, India}
\affiliation{Department of Physics, Ashoka University,  Sonipat, Haryana 131029, India}



\begin{abstract}
We report simultaneous detection of twin kHz and $\sim 40$ Hz quasi-periodic oscillations (QPOs) in the time-resolved analysis of the AstroSat/LAXPC observation of the neutron star low mass X-ray binary, 4U 1728-34. The frequencies of the multiple sets of triplets are correlated with each other and are consistent with their identification as the orbital, periastron and twice the nodal precessions frequencies. The observed relations, along with the known spin of the neutron star, put constraints on the mass and the ratio of moment of inertia to the mass of the neutron star to be  $M^*_\odot = 1.92\pm 0.01$ and $I_{45}/M^*_\odot = 1.07\pm 0.01$ under the simplistic assumption that the metric is a Kerr one. We crudely estimate that the mass and moment of inertia values obtained may differ by about 1\% and 5\%, respectively, if a self-consistent metric is invoked. Using the TOV equations for computing the moment of inertia of a neutron star in slow rotation approximation, having different equations of state, we find that the predicted values of neutron star parameters favor stiffer equations of state. We expect more stringent constraints would be obtained using a more detailed treatment, where the EOS-dependent metric is used to compute the expected frequencies rather than the Kerr metric used here. The results provide insight into both the nature of these QPOs and the neutron star interior. 

\end{abstract}

\keywords{Neutron star -- Black Hole -- X-ray binaries -- Quasi-Periodic Oscillations (QPOs) -- General Relativity (GR)}


\section{Introduction} \label{sec:intro}
The power density spectra (PDS) obtained from the X-ray light curves of X-ray binaries (XRBs) often reveal narrow features which are called quasi-periodic oscillations (QPOs) \citep{klis1989fourier,klis2000millisecond,ingram2019review}. QPOs, with frequencies ranging from a few mHz to around 1.2 kHz,  have been observed in different XRBs, which are broadly classified into mHz QPOs, low frequency (LF) QPOs (0.1 to 100 Hz), and high frequency (HF) QPOs above 300 Hz.
HF QPOs provide a useful probe into the inner accretion flows around the compact objects. The HF QPOs as well as  LF QPOs are observed in both black hole (BH)  and neutron star (NS) low-mass X-ray binaries (LMXBs). Understanding HF QPOs is expected to provide insights into the behavior of matter under strong gravitational ﬁeld. For NS systems, HF QPOs further provide the potential to constrain both mass \& radius and, hence, information about the equation of state (EOS) of the NS. The HF QPOs in NS-LMXBs are often observed as pairs, which are called lower \& upper kHz QPOs  \citep{van_der_Klis_1997}. 

NS-LMXBs are subdivided into Z sources and atoll sources on the basis of their evolution on the color-color diagram \citep{hasinger}. 
The first kHz QPOs  at $\sim$ 800 Hz and $\sim$ 1100 Hz were discovered in a Z source Sco-X1 from the observations made by the \textit{Rossi X-ray Timing Explorer (RXTE)} in 1996 \citep{van1996discovery, Berger_1996}. Apart from this discovery, \cite{van1996discovery} also reported correlations between the frequencies of the kHz and LF QPOs. Similar kHz QPO features have also been reported for atoll sources, and among these atoll sources, kHz QPOs were first discovered in 4U 1728-34 \citep{Strohmayer_1996}. In 4U 1728-34, \cite{Strohmayer_1996} reported three pairs of kHz QPOs in which the lower kHz QPO frequency increased from  $637.5\pm3.6$ Hz to $716.0\pm1.1$ Hz, whereas the upper kHz QPO frequency increased from $988\pm5.9$ Hz to $1058.3\pm12.1$ Hz. In all three PDS, the difference between lower and upper kHz QPOs frequencies was nearly equal to the burst oscillation (BO)  ($\sim$ 363 Hz) observed in the same observation, which is identified as the spin frequency of the NS. This led to the beat frequency model where the upper kHz QPO was identified as a keplerian frequency, while the lower one was the beat between the orbital frequency and spin frequency of the NS \citep{Alpar:1985jy,1985Natur.317..681L,Wijnands:2003ir}. However, more observations were in contradiction since the frequency difference turned out to be sometimes significantly smaller than the NS spin frequency \citep{Strohmayer_1996, Miller_1998,Psaltis_1998,Di_Salvo_2001}.

 \cite{Psaltis_1998} showed that the upper and lower kHz QPO frequencies are correlated as a power-law for Sco-X1, and a similar correlation was obtained when data from nine other sources were included. Moreover, correlations between the QPO frequencies and characteristic frequencies of the broad-band noise were discovered for both NS and BH systems  \citep{Psaltis_1999}. Thus, it was established that the frequencies of the different QPOs and broad-band noise were probably related to an underlying common process. 
 
These correlations led to the relativistic precession model (RPM), based on the motion of particles around a spinning BH \citep{Stella_1997,Stella_1999}.
In general relativity, the azimuthal frequency of a particle moving in a circular orbit around a Kerr black hole of mass $M$ and specific angular momentum  ($\tilde{a}=J/M)$ (where $J$ is the angular momentum), in an equatorial plane, is given by \citep{bardeen1972}
\begin{equation}\label{f11} 
\nu_\phi=\pm M^{1/2}r^{-3/2}\left[2\pi (1\pm \tilde{a}M^{1/2}r^{-3/2})\right]^{-1}.
\end{equation}
If such a particle's motion is perturbed along the polar angle  $\theta$ and radial distance $r$, then it would start oscillating in the directions of $\theta$ and $r$. The frequency of these small oscillations within the plane of the orbit ($\nu_r$) and perpendicular to the plane ($\nu_\theta$) are as follows \citep{okazaki1987global,kato1990trapped}
\begin{equation}\label{f12} 
\nu^2_r=\nu^2_\phi \left(1-6Mr^{-1}\pm 8\tilde{a}M^{1/2}r^{-3/2}-3\tilde{a}^2r^{-2}\right),
\end{equation}
\begin{equation}\label{f13} 
\nu^2_\theta=\nu^2_\phi \left(1\mp 4\tilde{a}M^{1/2}r^{-3/2}+3\tilde{a}^2r^{-2}\right),
\end{equation}
where $G$ and $c$ have been taken to be unity, and the positive and negative signs are for prograde and retrograde orbits, respectively.
These three coordinate frequencies give rise to two kinds of relativistic precessions: periastron and nodal precessions, and the corresponding precession frequencies are defined as $\nu_{per}=\nu_\phi -\nu_r$ and $\nu_{nod}=\nu_\phi -\nu_\theta$, respectively. According to the RPM, $\nu_\phi$, $\nu_{per}$ and  $\nu_{nod}$ are the frequencies of upper kHz, lower kHz  and low frequency QPOs, respectively.
An important aspect of the model is that the mass and spin parameter ($a=Jc/GM^2$) of the compact object can, in principle, be estimated using the correlations between the observed frequencies.

There have been several indications that the RPM is applicable to the QPO frequencies observed in NS systems \citep{Stella_1999,Ford_1998}.
\cite{Stella_1999} showed that correlations reported by \cite{Psaltis_1999}, where a large number of QPO frequencies from different systems were used, broadly follow the predictions of the model. The model has also been extended to BH systems. For two BH systems, the mass and spin parameter of the BHs have been estimated using observation of a QPO triplet for each case \citep{motta14,motta22}. In the absence of simultaneous observation of a QPO triplet, the correlation between the frequencies of two QPOs, consisting of a LF QPO and a HF QPO (or a broad-noise), has been used to estimate the BH spin parameter, using the BH mass estimated through optical measurements \citep{yash21,motta14b}. Detection of several simultaneous QPO triplets for a single source would allow us to verify the RPM and estimate the mass and spin parameter of the NS. If the spin is known independently, this will allow for an estimate of the star's moment of inertia and subsequently constrain the neutron star's EOS. 

In this work, we report the detection of three simultaneous QPOs in the power spectra of thirteen time segments as observed by AstroSat/LAXPC for the neutron star source 4U 1728-34. The data are unique because all thirteen segments contain QPO triplets corresponding to a single observation, showing a significant variation in QPO frequencies. We use the data to verify the RPM and constrain the mass and moment of inertia of the NS, which will eventually help to constrain the EOS of a
rotating NS. We also numerically compute the moment of inertia of NS spinning at 300 Hz for ten equations of state (EOSs). We take the EOS tables that are freely available and explore some by using them in our stellar models. The stellar model is constructed by solving the Tolman-Oppenheimer-Volkoff (TOV) equation with a numerical integration algorithm. This gives the stellar model of a
single star with a given core density and given EOS. The process is repeated for all
the EOSs discussed in this work. To get the plot of the moment of inertia vs. mass, we also take into account the constraints from the tidal deformability parameter and tidal love number.

\section{Observation and Data Analysis} \label{sec:style}
Large Area X-ray Proportional Counter (LAXPC) is an instrument onboard AstroSat satellite launched by the Indian Space Research Organization (ISRO) on the 28th of September 2015. LAXPC consists of co-aligned three proportional counter units
(PCUs) named LAXPC10, LAXPC20 and LAXPC30 with a large effective area of about 6000 cm$^2$, making it a perfect instrument to study even the faint X-ray sources. Each LAXPC detector records the arrival time of each photon with a time resolution
of 10 $\mu s$. In event analysis mode, each LAXPC instrument works in the energy range of 3-80 keV with an energy resolution of about 15 percent at 30 keV \citep{yadav2017large,Antia_2017}.\\
\begin{figure}[!h]\centering
	\includegraphics[width=0.8\linewidth]{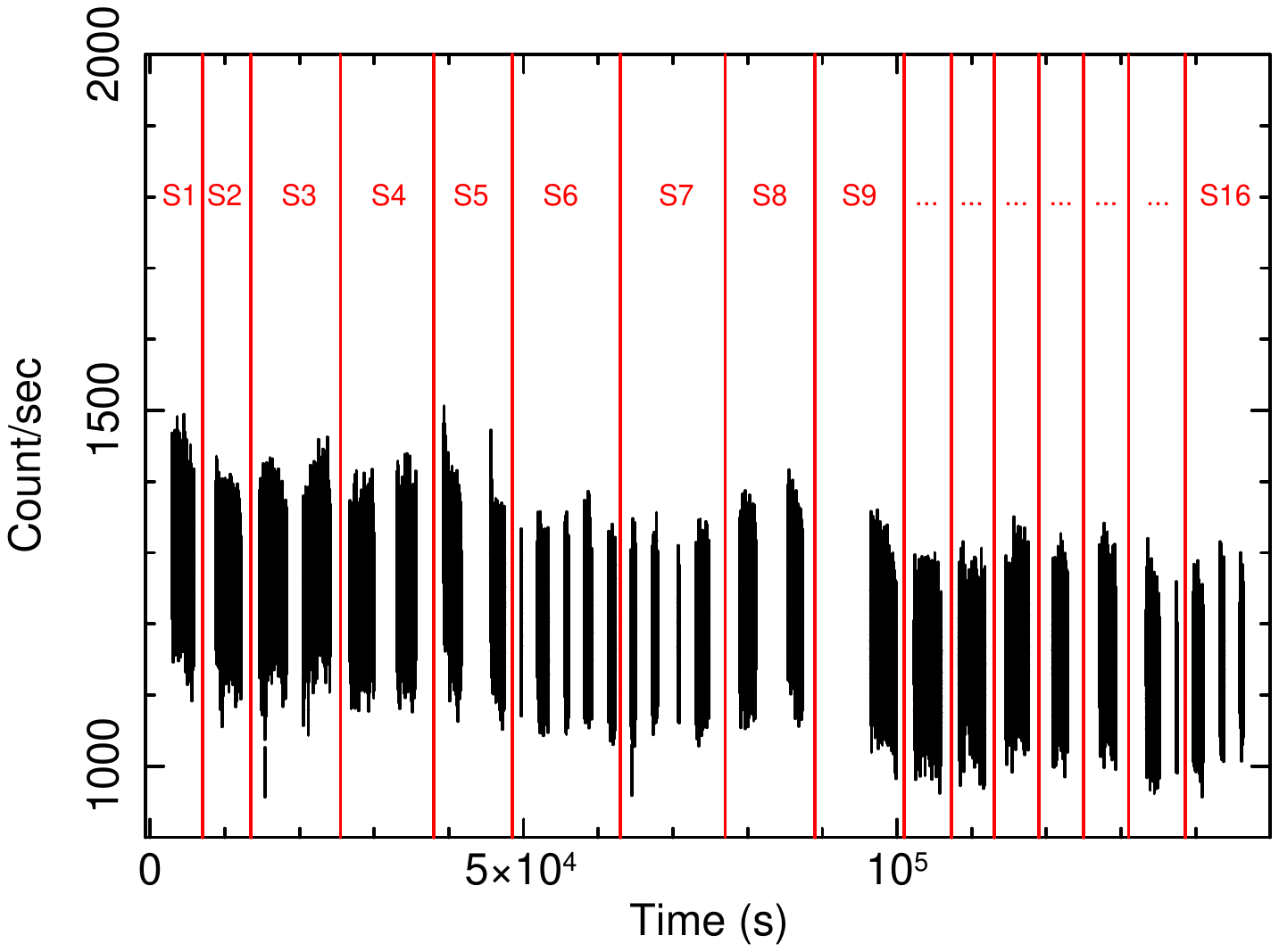}
	\caption{3-30 keV background-subtracted light curve of persistent emission after removing thermonuclear bursts. The light curve has been extracted from all layers of LAXPC10, LAXPC20, and LAXPC30 with a bin time bin of 1 s. The vertical red lines indicate the segments used to extract the PDS. S1, S2, S3, and so on S16 stand for segments 1, 2, 3, and so on segment 16, respectively.}
	\label{fig:lightcurve_1}
	
\end{figure}
In the course of its performance verification phase, AstroSat/LAXPC observed the source 4U 1728-34 covering 20 orbits during 7-8 March 2016 (observation ID: T01\_041T01\_9000000362)\footnote{\url{https://astrobrowse.issdc.gov.in/astro_archive/archive/Home.jsp}}. For one of these orbits (orbit number 2398) \citet{chauhan2017astrosat} reported a kHz QPO  at $\sim$ 800 Hz which showed evolution with time, burst oscillation at $\sim$ 363 Hz and performed spectral analysis of the persistent emission. In this work, we analyze the full data set. We used the \textit{LAXPCsoftware}\footnote{\url{http://astrosat-ssc.iucaa.in/laxpcData}} to compute the background subtracted light curves and power spectra. All three LAXPC units were used for the analysis. There are four thermonuclear bursts detected during this observations which were removed for this analysis. We defer a study of these bursts to a future work.

\begin{figure}\centering
	\includegraphics[width=\textwidth,angle=270, scale=0.55,trim={0 2.0cm 0 0}]{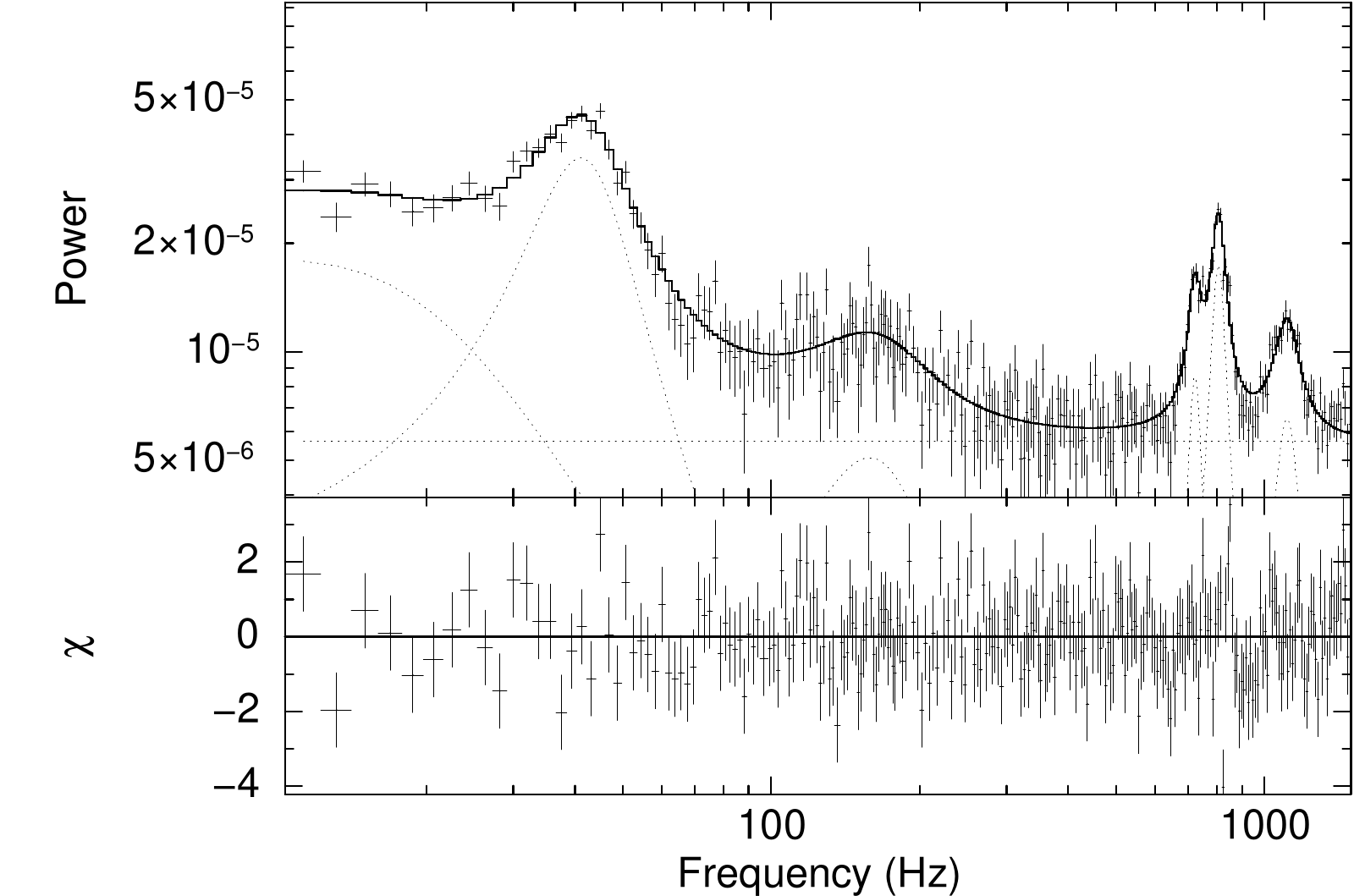}
	\caption{3-30 keV PDS of persistent emission in the frequency range from 10 Hz to 1500 Hz. It shows three clear QPOs at $\sim$ 40 Hz, $\sim$ 800 Hz, and $\sim$ 1100 Hz, along with a broad-band noise at $\sim$ 150 Hz. Lower kHz QPO shows double features at $\sim$ 723 Hz and $\sim$ 806 Hz.}
	\label{fig:pds_1}
\end{figure}

Figure~\ref{fig:lightcurve_1} shows the 3-30 keV background-subtracted light curve of persistent emission for the whole observation, while Figure~\ref{fig:pds_1} shows the corresponding power density spectrum (PDS). The software subtracts the expected dead-time corrected Poisson noise from the PDS. However, there was a small residual power at high frequencies, indicating that there is still some residual Poisson noise contribution. Hence, we included a constant represented by power-law with zero index to take into account this excess contribution. Six Lorentzian components were required to fit the time-averaged  power spectrum and their parameters, and the fit statistics are listed in  Table~\ref{tab:table_1}. At the lowest frequencies, there is a broad feature that is modeled as a Lorentzian with centroid 10 Hz (fixed) and width $\sim 33$ Hz. A prominent LF QPO is seen at $\sim  41$ Hz, along with a faint broad component at $\sim 160$ Hz. In the kHz region, complex shapes are seen, which are modeled by three relatively narrow Lorentzians. The complex shape, especially near 800 Hz, is due to varying QPO frequencies, and hence, the data has to be analyzed in segments to detect such variations.

\begin{figure}\centering
	\begin{minipage}{0.45\textwidth}
		\includegraphics[width=.6\linewidth,angle=270]{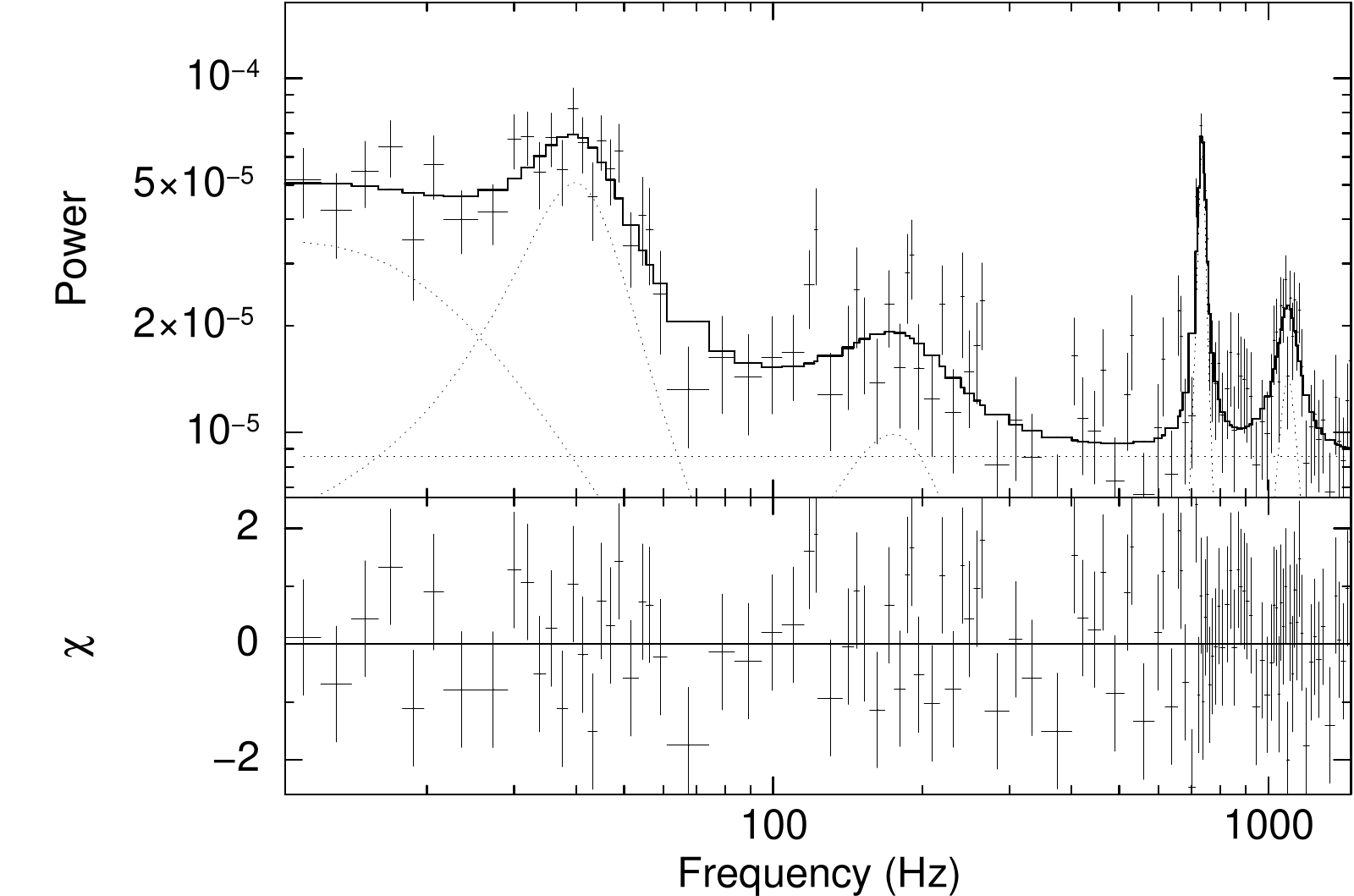}
		\caption{3-30 keV PDS of segment 6 in the frequency range from 10 Hz to 1500 Hz.}
		\label{fig:pds6}
	\end{minipage} \hspace{0.1cm}
	\begin{minipage}{0.45\textwidth}
		\includegraphics[width=.6\linewidth,angle=270]{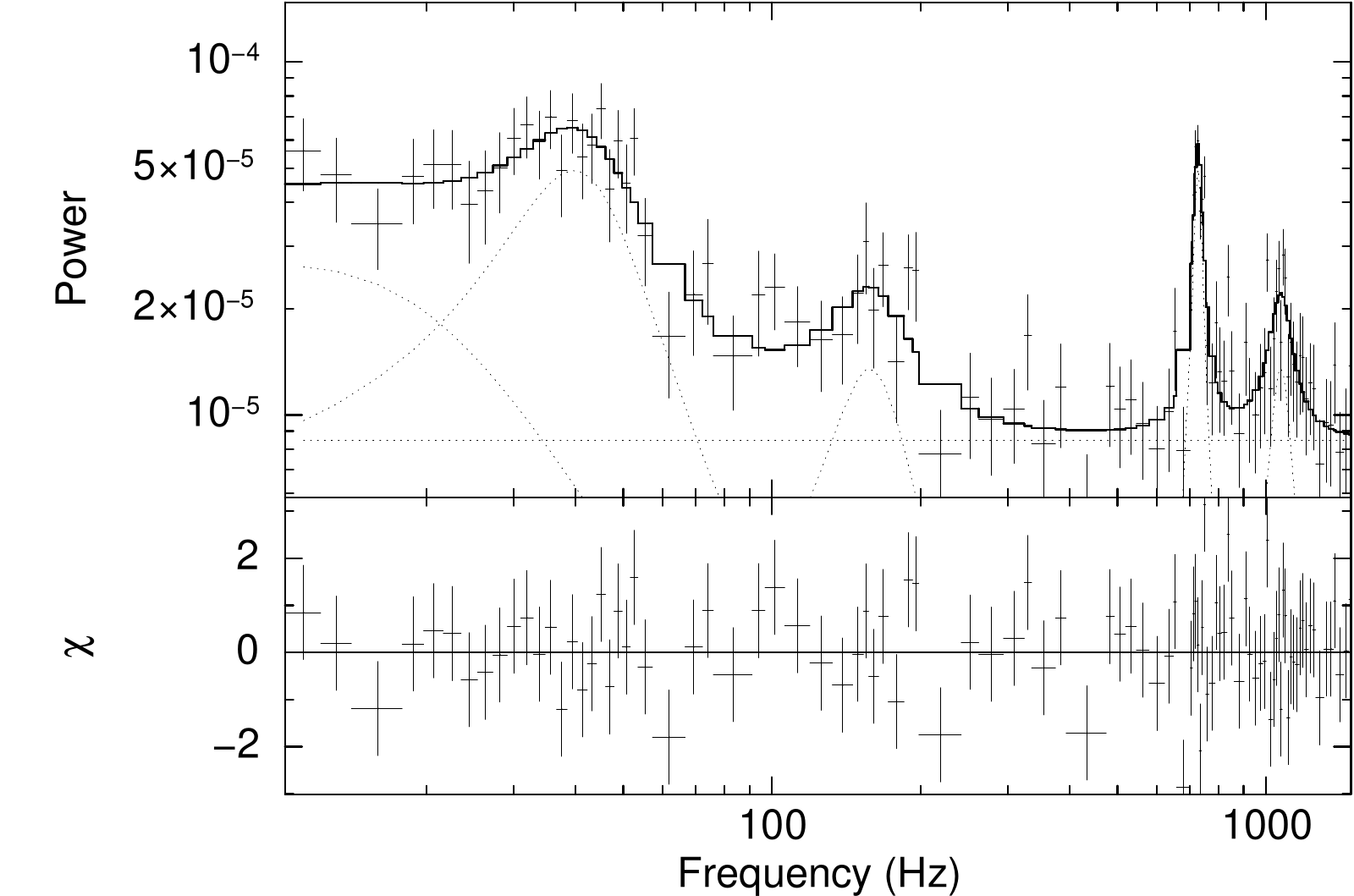}
		\caption{3-30 keV PDS of segment 7 in the frequency range from 10 Hz to 1500 Hz.}
		\label{fig:pds7}
	\end{minipage}
\end{figure}

To investigate the variation of the centroid frequencies of QPOs with time, we divided the observation into sixteen segments, as shown in Figure~\ref{fig:lightcurve_1}. The segments were chosen such that for the PDS generated the high-frequency features could be represented by either one or two Lorenztian(s) for the upper and lower kHz QPOs. Each individual PDS was fitted with a similar model used for the time-averaged one, except that centroid and width of the lowest frequency broad-band Lorentzian was not constrained and hence were fixed at 10 and 33.36 Hz, corresponding to the time-averaged values. 
The broad-band noise at $\sim$ 160 Hz generally had a low Q-factor ($\sim$ 1.4), and didn't appear in some of the segments. In this work, we concentrate on three distinct QPOs, namely the low frequency, lower, and upper kHz QPOs. Representative PDS of segments 6 and 7 and the fitting are shown in Figures 
\ref{fig:pds6} and \ref{fig:pds7}, and the best-fit parameters for these three QPOs are listed in Table~\ref{tab:table_2}. For three of the segments (1, 2, and 5), there was no detectable upper kHz QPO, and hence, there were 13 segments for which a triplet of QPOs was detected.

\begin{table}
	\centering
	\caption{The best-fit parameters of 3-30 keV PDS of persistent emission in 4U 1728-34 for entire observation. The errors quoted here are within 90 \% confidence level. The PDS has been fitted with six Lorentzians and one power-law with zero index. The reduced chi-squared obtained from the best fit is 1.22 for 244 degrees of freedom.}
	\label{tab:table_1}
	\begin{tabular}{llrr} 
		\hline
		\textbf{Model Component} & \textbf{Parameter (Unit)} & \textbf{Value}\\
		\hline\hline
		\textbf{1. \space lorentzian}  & \text{Peak Position (Hz)} & 10 (fixed) \\
		& \text{Width (Hz)}& $33.36_{-12.23}^{+21.28}$ & \\
		& \text{Norm ($10^{-4}$)}& $6.32_{-1.85}^{+3.32}$ & \\\\
		\textbf{2. \space lorentzian}  & \text{Peak Position (Hz)} & $41.12_{-0.86}^{+0.80}$ \\
		& \text{Width (Hz)}& $21.02_{-3.29}^{+2.97}$ & \\
		& \text{Norm ($10^{-4}$)}& $10.55_{-2.23}^{+1.72}$ & \\\\
		\textbf{3. \space lorentzian}  & \text{Peak Position (Hz)} & $158.30_{-9.06}^{+8.56}$ \\
		& \text{Width (Hz)}& $112.35_{-27.41}^{+35.32}$ & \\
		& \text{Norm ($10^{-4}$)}& $7.99_{-1.71}^{+1.81}$ & \\\\
		\textbf{4. \space lorentzian} & \text{Peak Position (Hz)} & $723.36_{-4.07}^{+3.90}$ \\
		& \text{Width (Hz)}& $48.94_{-10.84}^{+11.99}$ & \\
		& \text{Norm ($10^{-4}$)}& $6.62_{-1.33}^{+1.39}$ & \\\\
		\textbf{5. \space lorentzian}  & \text{Peak Position (Hz)} & $806.75_{-2.12}^{+1.87}$ \\
		& \text{Width (Hz)}& $60.56_{-6.33}^{+7.39}$ & \\
		& \text{Norm ($10^{-4}$)}& $16.52_{-1.49}^{+1.68}$ & \\\\
		\textbf{6. \space lorentzian}  & \text{Peak Position (Hz)} & $1110.95_{-7.51}^{+7.54}$ \\
		& \text{Width (Hz)}& $146.99_{-21.69}^{+25.21}$ & \\
		& \text{Norm ($10^{-4}$)}& $14.79_{-2.02}^{+2.26}$ & \\\\
		\textbf{7. \space powerlaw}  & \text{Photon Index} & 0 (fixed) \\
		& \text{Norm ($10^{-6}$)}& $5.64_{-0.28}^{+0.26}$ & \\ \\
		& $\chi^2$/dof &  298.12/244 \\
		\hline
	\end{tabular}
\end{table}

\begin{deluxetable*}{lllcllcllcr}
		\caption{The best-fit parameters of lower and upper kHz QPOs observed in 4U 1728-34 for each segment . The errors quoted here are within $1 \sigma$ confidence level.}
	\label{tab:table_2}
	\tablewidth{700pt}
	\tabletypesize{\scriptsize}
	\tablehead{
		\colhead{Segment} &  \multicolumn{3}{c}{LF QPO}  &  \multicolumn{3}{c}{Lower kHz QPO}  &  \multicolumn{3}{c}{Upper kHz QPO}  &  ${\chi^{2}/dof}$\\
		 \colhead{Number} &  \colhead{Position} & \colhead{Width} & \colhead{Norm}  &  \colhead{Position} & \colhead{Width} & \colhead{Norm}  & \colhead{Position} & \colhead{Width} & \colhead{Norm}  &  \colhead{}\\
		  \colhead{} &  \colhead{(Hz)} & \colhead{(Hz)} & ${(10^{-4})}$  &  \colhead{(Hz)} &
		 \colhead{(Hz)} & ${(10^{-4})}$  & \colhead{(Hz)} & \colhead{(Hz)} & ${(10^{-4})}$  &  \text{} \\
	} 
	\startdata\\
	\text{1} & $44.26_{-3.51}^{+2.19}$ & $23.71_{-9.73}^{+13.91}$ & $9.55_{-2.73}^{+3.68}$ & $852.65_{-0.53}^{+0.55}$ & $14.88_{-1.43}^{+1.25}$ & $28.46_{-1.67}^{+1.46}$ & - & -  & - & 72.44/48\\\\
\text{2} & $42.68_{-1.23}^{+1.29}$ & $18.02_{-3.40}^{+4.45}$ & $11.47_{-1.72}^{+2.05}$ & $802.61_{-0.69}^{+0.59}$ & $13.38_{-0.78}^{+1.90}$ & $25.33_{-1.21}^{+1.87}$ & - & - & - & 96.90/64\\\\
\text{3} & $42.59_{-1.07}^{+1.08}$ & $26.60_{-2.72}^{+3.18}$ & $15.37_{-0.92}^{+0.96}$ & $812.02_{-0.61}^{+0.64}$ & $25.57_{-1.25}^{+1.55}$ & $32.48_{-1.05}^{+1.04}$ & $1135.95_{-10.96}^{+11.02}$ & $73.43_{-13.57}^{+17.23}$ & $9.84_{-1.83}^{+1.81}$ & 148.19/114\\\\
\text{4} & $42.19_{-1.05}^{+0.99}$ & $18.68_{-3.03}^{+3.69}$ & $12.55_{-1.61}^{+1.79}$ & $811.04_{-2.32}^{+2.06}$ & $58.03_{-4.88}^{+5.17}$ & $35.19_{-2.52}^{+2.61}$ & $1113.94_{-15.86}^{+19.39}$ & $168.03_{-57.33}^{+68.78}$ & $22.19_{-6.38}^{+7.84}$ & 129.63/96\\\\
\text{5} & $44.58_{-1.02}^{+1.01}$ & $14.66_{-2.30}^{+2.74}$ & $10.83_{-1.38}^{+1.49}$ & $823.21_{-3.17}^{+2.82}$ & $67.66_{-5.99}^{+6.40}$ & $39.03_{-2.93}^{+3.01}$ & - & - & - & 105.92/76\\\\
\text{6} & $38.60_{-1.74}^{+1.69}$ & $25.59_{-4.08}^{+4.97}$ & $17.65_{-2.62}^{+3.03}$ & $734.94_{-1.32}^{+1.15}$ & $29.08_{-3.34}^{+3.68}$ & $26.21_{-2.28}^{+2.39}$ & $1098.42_{-10.79}^{+11.24}$ & $154.35_{-45.09}^{+66.31}$ & $31.62_{-7.78}^{+10.84}$ & 127.50/102\\\\
\text{7} & $41.08_{-2.07}^{+1.86}$ & $27.13_{-5.27}^{+6.71}$ & $18.39_{-3.02}^{+3.59}$ & $728.41_{-2.62}^{+1.38}$ & $32.03_{-5.22}^{+4.46}$ & $24.49_{-2.85}^{+2.65}$ & $1074.21_{-9.48}^{+8.98}$ & $107.73_{-29.78}^{+38.25}$ & $25.81_{-5.78}^{+6.73}$ & 109.98/87\\\\
\text{8} & $43.07_{-0.94}^{+1.01}$ & $18.19_{-3.70}^{+4.65}$ & $13.99_{-2.05}^{+2.35}$ & $785.46_{-1.49}^{+0.87}$ & $30.03_{-3.45}^{+3.35}$ & $28.47_{-2.29}^{+2.35}$ & $1138.49_{-14.03}^{+12.76}$ & $106.32_{-35.74}^{+46.30}$ & $15.87_{-4.37}^{+5.02}$ & 78.85/76\\\\
\text{9} & $39.03_{-2.37}^{+2.01}$ & $29.25_{-5.58}^{+7.12}$ & $20.14_{-3.46}^{+4.30}$ & $730.55_{-2.71}^{+2.64}$ & $54.98_{-6.40}^{+7.25}$ & $35.03_{-3.41}^{+3.63}$ & $1066.39_{-14.58}^{+14.39}$ & $178.81_{-51.46}^{+75.05}$ & $33.78_{-8.76}^{+3.63}$ & 66.72/85\\\\
\text{10} & $34.59_{-1.05}^{+1.05}$ & $17.07_{-3.11}^{+3.79}$ & $16.24_{-2.91}^{+3.20}$ & $674.37_{-1.43}^{+1.09}$ & $22.16_{-3.85}^{+4.51}$ & $18.80_{-2.27}^{+2.39}$ & $1036.10_{-5.97}^{+6.13}$ & $93.58_{-20.25}^{+26.09}$ & $31.08_{-5.31}^{+6.22}$ & 97.85/98\\\\
\text{11} & $39.19_{-2.04}^{+2.15}$ & $12.47_{-5.15}^{+7.76}$ & $5.90_{-2.29}^{+3.33}$ & $710.02_{-1.44}^{+1.44}$ & $26.34_{-3.39}^{+3.82}$ & $25.86_{-2.57}^{+2.72}$ & $1071.45_{-17.69}^{+29.85}$ & $120.08_{-55.63}^{+94.10}$ & $19.71_{-7.38}^{+10.94}$ & 68.21/70\\\\
\text{12} & $41.56_{-1.49}^{+1.43}$ & $18.35_{-3.56}^{+4.19}$ & $13.79_{-2.27}^{+2.45}$ & $760.83_{-1.04}^{+0.88}$ & $21.49_{-2.15}^{+2.29}$ & $28.81_{-2.22}^{+2.26}$ & $1107.30_{-8.42}^{+7.16}$ & $63.36_{-8.42}^{+7.16}$ & $17.65_{-4.28}^{+5.24}$ & 83.07/63\\\\
\text{13} & $39.91_{-2.08}^{+2.15}$ & $22.75_{-6.87}^{+10.42}$ & $14.89_{-3.71}^{+4.80}$ & $713.74_{-1.38}^{+0.73}$ & $19.31_{-2.82}^{+2.96}$ & $26.95_{-2.81}^{+2.81}$ & $1086.65_{-8.76}^{+8.75}$ & $85.02_{-22.44}^{+30.15}$ & $26.29_{-6.06}^{+7.16}$ & 60.29/60\\\\
\text{14} & $44.02_{-0.98}^{+0.99}$ & $11.51_{-2.52}^{+3.07}$ & $11.12_{-2.24}^{+2.16}$ & $763.27_{-0.36}^{+0.33}$ & $8.99_{-1.94}^{+2.04}$ & $25.89_{-2.09}^{+2.16}$ & $1114.83_{-17.04}^{+18.25}$ & $123.41_{-41.67}^{+66.27}$ & $21.86_{-6.50}^{+8.36}$ & 46.32/50\\\\
\text{15} & $36.75_{-1.57}^{+1.44}$ & $18.62_{-3.63}^{+4.46}$ & $19.76_{-3.38}^{+3.93}$ & $701.18_{-1.98}^{+2.04}$ & $28.94_{-5.75}^{+6.94}$ & $23.89_{-3.54}^{+3.85}$ & $1060.47_{-12.67}^{+14.38}$ & $114.00_{-37.43}^{+47.91}$ & $33.92_{-8.98}^{+11.14}$ & 61.62/65\\\\
\text{16} & $39.69_{-1.49}^{+1.37}$ & $21.82_{-4.92}^{+6.46}$ & $19.85_{-3.78}^{+4.56}$ & $729.87_{-5.21}^{+6.46}$ & $71.78_{-21.68}^{+29.59}$ & $30.84_{-6.75}^{+8.19}$ & $1096.50_{-27.51}^{+22.19}$ & $230.78_{-75.08}^{+115.23}$ & $41.58_{-12.84}^{+18.39}$ & 64.19/63\\\\
\enddata
\end{deluxetable*}

\section{Results} \label{sec:results}
As mentioned earlier, following the relativistic precession model, $\nu_\phi$, $\nu_{per}$ and  $\nu_{nod}$ are identified with the frequencies of upper kHz, lower kHz  and LF QPOs, respectively. It is convenient to express $\nu_{per}$ and $\nu_{nod}$ is terms of $\nu_\phi$, by eliminating the radius $r$ from Equations 1, 2, and 3. Introducing the constants $G$ and $c$ into the expressions and noting that the specific angular momentum can be written as $\tilde a =\frac{2\pi I\nu_s}{M} $ , where  $\nu_{s}$ is the spin frequency, one can write $r$ as a function of $\nu_\phi$
\begin{equation}
r(\nu_\phi)=\pm (GM)^{1/3}\left[\frac{1}{2\pi \nu_\phi}-\left(\frac{I}{M}\right)\frac{2\pi \nu_{s}}{c^2}\right]^{2/3}
\end{equation}
where $M$ and $I/M$ are mass and ratio of moment of inertia to the mass, respectively. The expressions for $\nu_{per}$ and $\nu_{nod}$ to have $M$ and $I/M$ as free parameters can be written as 
\begin{equation}
\nu_{per}=\nu_\phi\left[1-\sqrt{1-\frac{6GM}{c^2r}\pm \frac{16\pi \nu_{s}(GM)^{1/2}}{c^2r^{3/2}}\left(\frac{I}{M}\right)-\frac{12\pi^2\nu_{s}^2}{r^2c^2}\left(\frac{I}{M}\right)^2}\right]
\end{equation}
\begin{equation}
\nu_{nod}=\nu_\phi\left[1-\sqrt{1\mp\frac{8\pi \nu_{s}(GM)^{1/2}}{c^2r^{3/2}}\left(\frac{I}{M}\right)+\frac{12\pi^2\nu_{s}^2}{r^2c^2}\left(\frac{I}{M}\right)^2}\right].
\end{equation}

\begin{figure}\centering
	\includegraphics[width=0.75\textwidth, scale=0.5]{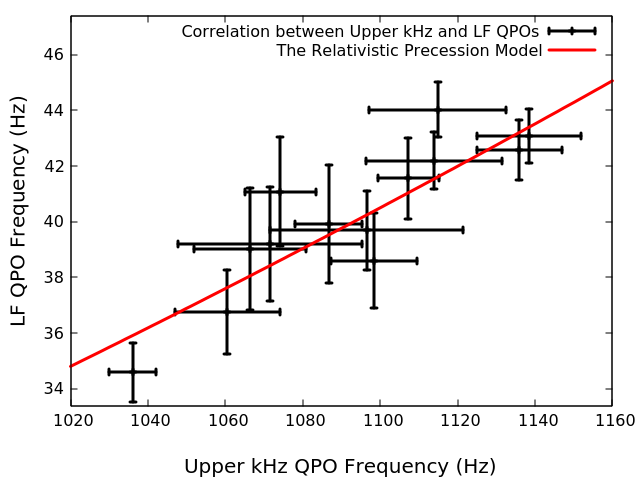}
	\caption{Correlation between LF QPO and upper kHz QPO fitted with the RPM. The best-fit parameter values obtained are $a=0.269\pm0.01$ (equivalently $I_{45}/M^*_\odot = 2.21\pm 0.02$) and $2.12$ (fixed) with $\chi_{red}^{2}=0.57$ for 12 dof.}
	\label{fig:corr1}
\end{figure}

\begin{figure}\centering
	\includegraphics[width=0.75\textwidth, scale=0.5]{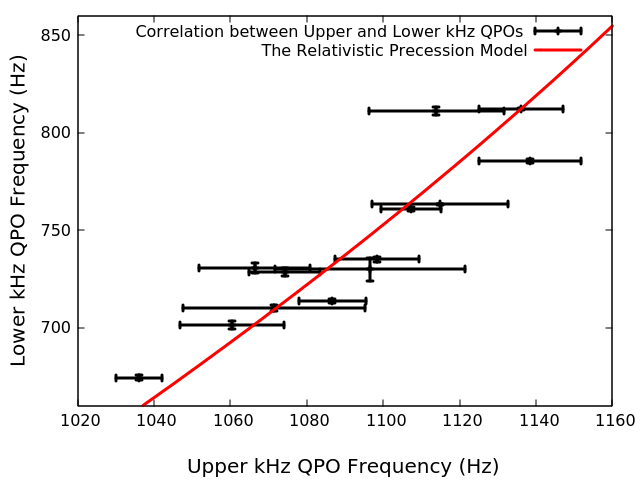}
	\caption{Correlation between lower and upper kHz QPOs. The best-fit parameter values obtained are $a=0.269$ (fixed) (equivalently $I_{45}/M^*_\odot = 2.21$) and $2.12\pm0.01$ with $\chi_{red}^{2}=1.06$ for 12 dof. Together with Figure~\ref{fig:corr1}, it shows the consistency of the RMP with QPO triplet.}
	\label{fig:corr2}
\end{figure}

\begin{figure}\centering
	\includegraphics[width=0.75\textwidth, scale=0.5]{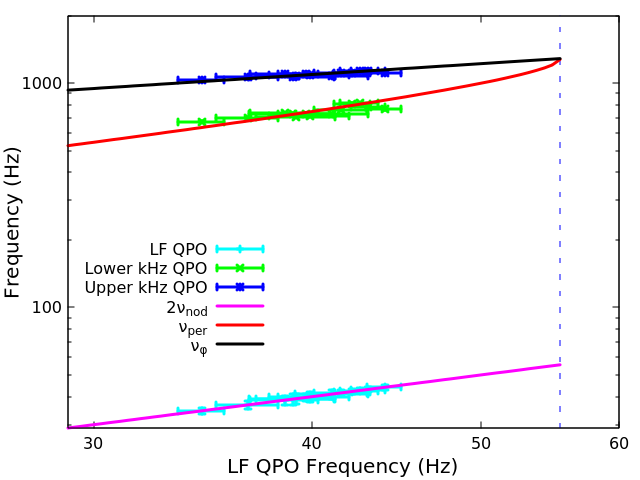}
	\caption{LF QPO, lower and upper kHz QPOs are plotted as a function of LF QPO. The solid lines represent characteristic frequencies of the RPM, corresponding to $M^*_\odot= 1.92$ and $I_{45}/M^*_\odot= 1.07$. The orbital frequency, the periastron precession frequency, and twice the nodal precession frequency are represented by solid ``black", ``red", and ``magenta" lines, respectively. The vertical dashed-line depicts the maximum QPO frequencies at the inner-most stable circular orbit.}
	\label{fig:rpm}
\end{figure}

While fitting the observed QPO frequencies with the RPM, we have taken $I$ in units of $10^{45}$ g cm$^2$ ($I_{45}$) and $M$ in solar mass ($M^*_\odot$). Moreover, we have considered the motion of the particle in a prograde orbit. These expressions can then be compared with the observed correlations between the frequencies of the lower kHz and LF QPOs with respect to the upper kHz QPO. Figures \ref{fig:corr1} and \ref{fig:corr2} show these observed correlations. The best-fit values obtained are $M^*_\odot= 2.12 \pm 0.01$ and $I_{45}/M^*_\odot= 2.21 \pm0.02 $. As shown in the next section, this value for 
$I_{45}/M^*_\odot$, is significantly larger than that expected for NSs based on theoretical studies on their EOS. One possibility is that the  LF QPO frequency may be equal to twice the nodal precession frequency. Indeed, some collective observational data points for a few atoll-type sources \citep{Morsink_1999,Stella_1999} suggest that this is the case, and  \cite{Stella_1999} have argued that the geometry of tilted inner accretion disk might produce a stronger signal at the second harmonic of the nodal precession frequency. Thus, we also considered the situation where the LF QPO frequency is twice the nodal precession frequency, which resulted in $M^*_\odot= 1.92\pm0.01$ and $I_{45}/M^*_\odot= 1.07\pm0.01$ (equivalently $a=0.145\pm0.001$). Figure~\ref{fig:rpm} shows the plot of LF QPO, lower kHz QPO, and upper kHz QPO frequencies as a function of LF QPO frequency. Solid lines represent orbital, periastron precession, and twice the nodal precession frequencies corresponding to the best-fit values of $M^*_\odot= 1.92$ and $a=0.145$. The vertical dashed-line indicates the maximum QPO frequencies at $r_{ISCO}=5.52\,r_g$. The Figure may be compared to the ones shown in the literature for BH systems \citep{motta14, motta22}, where the theoretical curves were plotted against a single triplet QPO observation.

While using Equations (1), (2), and (3), we have assumed that the metric to be a Kerr one, while the actual metric around a spinning NS will be different and will depend on its EOS \citep{Morsink_1999}. Since such a metric has to be generally obtained numerically, we defer a detailed analysis to a later work. Here, we attempt to estimate the effect of the Kerr metric assumption by ignoring the second order $\tilde{a}^2$ terms in the equation and fitting the correlations to obtain  $M^*_\odot= 1.93\pm0.01$ and $I_{45}/M^*_\odot= 1.03\pm0.01$. The difference in the parameters obtained, which is $<$  1\% for $M^*_\odot$ and about 5\% for $I_{45}/M^*_\odot$, may reflect the deviation caused by assuming a Kerr metric. Thus, considering these systematic effects, we estimate the approximate values to be $M^*_\odot= 1.92\pm0.03$ and $I_{45}/M^*_\odot= 1.07\pm0.05$. The systematics estimated by ignoring the second order of spin is a crude estimation. In fact, the metric around an NS for a given EOS would depend on a number of parameters such as stellar oblateness, higher-order mass moments, magnetic field, etc. Hence, there may exist a significant degree of uncertainty associated with these systematics.
\section{Constraining Equation of State}\label{MoI}

In this section, we make an approximate estimate of the moment of inertia of a NS, given an equation of state, i.e. pressure as a function of density $P(\rho)$. In particular, we numerically solve Tolman-Oppenheimer-Volkoff (TOV) equations for an EOS to obtain pressure as a function of radius $P(r)$. We note that the TOV equations assume that the underlying metric is a Schwarschild one. The moment of inertia is estimated numerically using the equation (11) mentioned in \cite{RavandP}, given by:

\begin{equation}\label{RavP}
I = \frac{J}{1 + \frac{2GJ}{R^3 c^2}}\ \ \ ; \ \ \  J \equiv \frac{8}{3}\pi\int_0^R r^4 \left(\rho(r) + \frac{P(r)}{c^2}\right) \alpha(r) \text{d}r
\end{equation}	
where $\alpha(r) = \left(1-\frac{2GM(r)}{Rc^2} \right)^{-1}$. \\
Here it is assumed that the neutron star is slowly rotating corresponding to a spin of $\sim 300$ Hz.

\begin{figure}
	\centering
	\includegraphics[scale=0.5]{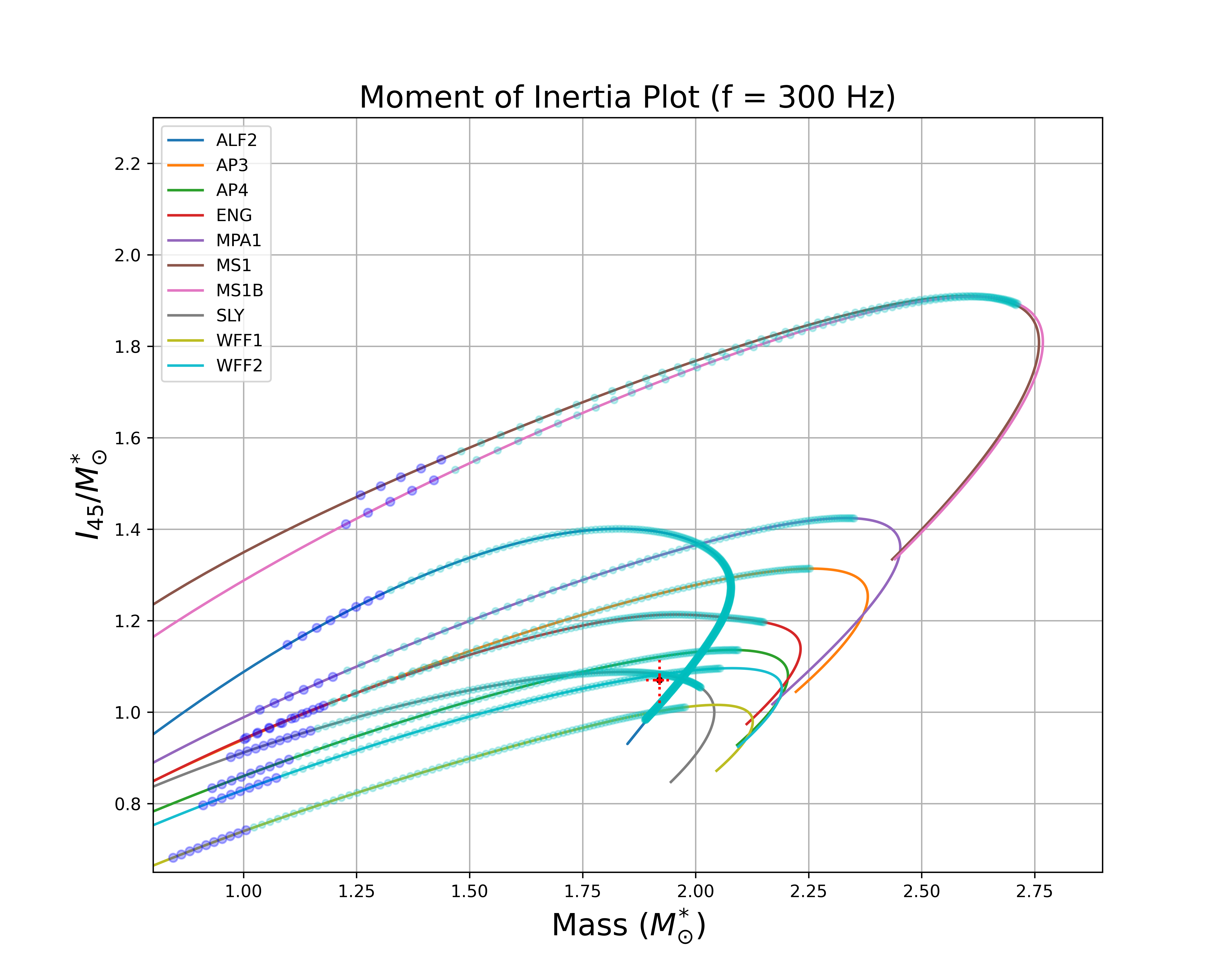}
	\caption{In the above plot the solid lines represent the variation of moment of
		inertia  ($I$) with the mass of the neutron star  ($M$) for different equations of
		state. The cyan coloured dots represent the GW170817 constraints for the 50\%
		confidence interval and blue dots represent the GW170817 constraints for the 90\%
		confidence interval. The red dot represents the observed values of $I$ and $M$ from the QPOs correlation with the RPM,
		and red bar (solid/dotted) denotes the statistical/systematic error in
		measurement. The equations of state considered here are as follows:
		ALF2\citep{alford2005hybrid}, AP3\citep{akmal1998equation}, AP4\citep{akmal1998equation}, ENG\citep{engvik1995asymmetric}, MPA1\citep{muther1987nuclear},
		MS1\citep{mueller1996relativistic}, MS1B\citep{mueller1996relativistic}, SLY\citep{douchin2001unified}, WFF1\citep{wiringa1988equation}, WFF2\citep{wiringa1988equation}.}
	\label{fig:MoI}
\end{figure}

\begin{figure}
	\centering
	\includegraphics[scale=0.5]{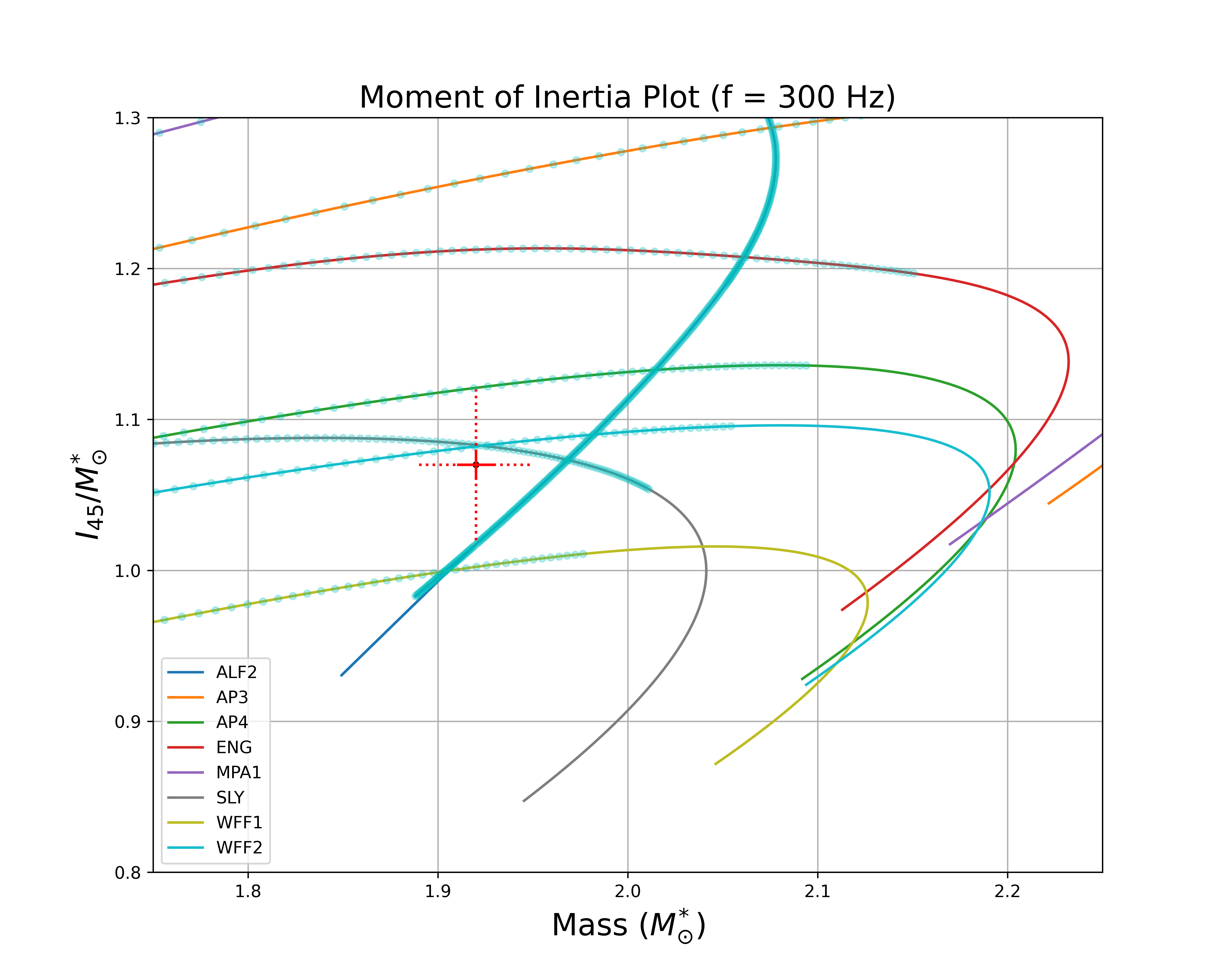}
	\caption{The above plot is a zoomed-in version of the Figure~\ref{fig:MoI}. Red cross in the plot indicates the location of the parameter space constrained by this work. }
	\label{fig:Zoomed}
\end{figure}

Figure~\ref{fig:MoI}  shows the results of the calculations with solid lines representing the variation of $I_{45}/M^*_\odot$ with $M^*_\odot$ for different EOSs mentioned in the caption. Note that for all the EOSs considered,  $I_{45}/M^*_\odot < 2$.
The Gravitational waveform for the event  GW170817 leads to constraints on the dimensionless tidal deformability number, $\Lambda$ \citep{abbott2018gw170817}. The cyan (blue) colored points in Figure~\ref{fig:MoI}, represent the 50\% (90\%) confidence levels obtained from these  $\Lambda$ constraints, which we have obtained by solving the relevant equations \citep{thorne1967non,hinderer2008tidal,hinderer2010tidal,chatziioannou2020neutron}. The constraints on M and I/M obtained from this work are marked in the Figure~\ref{fig:MoI} with a red point with solid and dotted bars representing the statistical and systematic errors, respectively. Figure~\ref{fig:Zoomed} shows a zoomed-up version of Figure~\ref{fig:MoI}, which reveals the location of the parameter space favored by the RPM fitting to the AstroSat data.

\section{Summary and Discussion}
To study kHz QPOs, we have thoroughly analyzed twenty orbits data of 4U 1728-34 obtained from the observation made by AstroSat/LAXPC during 7-8 March 2016. After removing thermonuclear bursts, we extracted 3-30 keV PDS of persistent emission for the whole observation (Figure~\ref{fig:lightcurve_1}). For PDS extraction, we used all layers of three LAXPCs and kept the frequency bin at 2 Hz. We have fitted the PDS for the entire data set with six Lorentzians and one power law with zero spectral index to account for some Poisson noise residuals at higher frequencies. The time-averaged PDS clearly shows LF QPO at $\sim$ 41 Hz and twin kHz QPOs at $\sim$ 800 Hz and $\sim$ 1100 Hz, along with a broad feature at $\sim$ 158 Hz (Figure~\ref{fig:pds_1} \& Table~\ref{tab:table_1}). A complex feature near 800 Hz showing a double-peaked lower kHz QPO was detected in the PDS. This complex feature disappeared once the data were divided into segments, which was indicative of significant variation in lower kHz QPO for the whole observation.

To see the time evolution of QPOs, we divided the entire data set into sixteen segments. For each segment, we computed 3-30 keV PDS with a 2 Hz frequency bin. QPO frequencies were found to vary with segment number, manifesting a dynamic nature of QPOs. LF QPO, lower and upper kHz QPOs frequencies varied from $\sim$ 35 Hz to 45 Hz, $\sim$ 674 Hz to 853 Hz, and $\sim$ 1036 Hz to 1136 Hz, respectively (Table~\ref{tab:table_2}). For three out of sixteen segments, upper kHz QPO was not detected. Thus, we could get thirteen QPO triplets, making it the first-ever detection of multiple sets of QPO triplets from a single observation. 

We used the Relativistic Precession Model to fit the observed QPO frequencies. As discussed in the previous section, in the RPM, we assumed LF QPO frequency to be twice the nodal precession frequency to estimate the mass and moment of inertia of NS by using BO frequency as spin frequency. This leads to the best-fit estimates $M^*_\odot= 1.92\pm0.01$ and $I_{45}/M^*_\odot= 1.07\pm0.01$. We used the RPM to fit the observed QPO frequencies, although the metric around a spinning NS deviates from that of the Kerr metric and depends upon the stellar structure and, hence, EOS. The exact metric around a rotating NS depends on the EOS and has to be computed
numerically. Hence, we postpone a detailed fit with a more realistic metric for future work. Here, to estimate the effect of the Kerr metric assumption, we ignored the terms involving second order of $\tilde{a}$ in the equations (equations \ref{f12} \& \ref{f13}) and obtained values of  $M^*_\odot$ and $I_{45}/M^*_\odot$ which differ by about 1 \% and 5 \%, respectively. If we consider these differences as systematic errors due to the Kerr metric assumptions, then we obtain $M^*_\odot= 1.92\pm0.03$ and $I_{45}/M^*_\odot= 1.07\pm0.05$ for the NS in 4U 1728-34. 

It is interesting to note that the LF QPO in 4U 1728-34 is sometimes observed to have harmonics. \cite{motta17} noted that in the integrated fractional RMS versus centroid frequency plot, both the harmonic and fundamental frequencies follow two nearly parallel tracks, indicative of a harmonic relationship. Thus, when the harmonics are not observed, it is possible that the single observed QPO actually corresponds to the second harmonic and not the fundamental. This would have provided a natural explanation as to why the LF QPO reported in this work needs to be identified with twice the nodal precession frequency. However, \cite{motta17} point out that the branch below $13\, \%$ RMS corresponds to the fundamental frequency, while the branch above $13\, \%$ corresponds to the second harmonic. The total RMS of the PDS in the frequency range of 0.1-64 Hz and energy range 3-15 keV (same as used by \cite{motta17}), reported in this work is $\sim 5\, \%$, and hence, it seems the LF QPO is the fundamental one and not the second harmonic. Thus, the identification of LF QPO as twice the nodal precession frequency has to be associated with dynamic nature of the system, as mentioned by \cite{Stella_1999}.

We estimated the moment of inertia of a NS for ten different EOSs using the TOV equations and found that our results favor relatively stiff EOS (WFF2, SLY, AP4, and ALF2) and are consistent with constraints obtained from the Gravitational-wave event GW170817. However, we note that the TOV equation assumes a Schwarschild metric, which would not be valid for a spinning NS. The statistical error on the estimates of the NS mass and moment of inertia is small ($<$ 1\%), and hence, a more sophisticated analysis where the test particle frequencies and the moment of inertia are computed for the correct metric for a given EOS, will lead to unprecedented constraints on the EOS. However, there will still be uncertainties arising from the unknown physical origin of the QPOs and, in particular, whether realistic hydro-dynamical flows will exhibit the same frequencies as computed for test particles. Nevertheless, these observations of several triplets of QPOs for a single source, with a known spin period, provides insight into the nature of the QPOs and holds the promise for stringent constraints on the EOS of neutron stars.
\\\\ \\ \\

\section*{Acknowledgements}
The authors express their gratitude to the anonymous reviewer for the insightful remarks and very constructive suggestions. This project has used the data of ISRO's satellite AstroSat/LAXPC. The data analyses have been carried out using LAXPCsoftware and HeaSoft of NASA's High Energy Astrophysics Science Archive Research Centre (HEASARC). We acknowledge the support from Indian Space Science Data Centre (ISSDC) as well as LAXPC POC at TIFR. K Anand expresses his gratitude to IUCAA for providing him with local hospitality during the completion phase of this work. We also acknowledge the Ashoka University HPC resources (Chanakya@Ashoka) made available for conducting a part of the research reported in this paper. 
\vspace{2cm}
%
\bibliography{sample631}{}
\bibliographystyle{aasjournal}


\end{document}